\documentclass[english,12pt]{article}
\usepackage{array}
\usepackage{graphicx}
\usepackage{amssymb}
\usepackage[english]{babel}
\usepackage{amsmath}
\usepackage{multirow}
\usepackage{prettyref}
\usepackage{babel}
\usepackage{units}
\usepackage[latin1]{inputenc}
\usepackage{amsfonts}
\usepackage{array}
\usepackage{babel}
\usepackage{color}
\usepackage{cite}
\def\@fmsl@sh#1#2#3{\m@th\ooalign{$\hfil#1\mkern#2/\hfil$\crcr$#1#3$}}
 \def\eq#1\en{\begin{equation}#1\end{equation}}
\def\s[#1,#2]{[#1\stackrel{\star}{,}#2]}
\def\sx[#1,#2]{[#1\stackrel{\star_{x}}{,}#2]}

\newcommand{\nc}{\newcommand}
\nc{\beq}{\begin{equation}}
\nc{\eeq}{\end{equation}}
\nc{\beqa}{\begin{eqnarray}}
\nc{\eeqa}{\end{eqnarray}}

\def\bc{\begin{center}}
\def\ec{\end{center}}

\def\to{\rightarrow}

\def\gsim{\mathrel{\mathpalette\atversim>}}

\def\bc{\begin{center}}
\def\ec{\end{center}}

\def\gsim{\mathrel{\rlap{\lower4pt\hbox{\hskip1pt$\sim$}}

    \raise1pt\hbox{$>$}}}       

\def\gsim{\mathrel{\rlap{\lower4pt\hbox{\hskip1pt$\sim$}}
    \raise1pt\hbox{$>$}}}       

\begin{document}
\makeatletter
\def\fmslash{\@ifnextchar[{\fmsl@sh}{\fmsl@sh[0mu]}}
\def\fmsl@sh[#1]#2{%
  \mathchoice
    {\@fmsl@sh\displaystyle{#1}{#2}}%
    {\@fmsl@sh\textstyle{#1}{#2}}%
    {\@fmsl@sh\scriptstyle{#1}{#2}}%
    {\@fmsl@sh\scriptscriptstyle{#1}{#2}}}
\def\@fmsl@sh#1#2#3{\m@th\ooalign{$\hfil#1\mkern#2/\hfil$\crcr$#1#3$}}
\makeatother

\thispagestyle{empty}
\begin{titlepage}
\boldmath
\begin{center}
  \Large {\bf  Hidden Sector and Gravity}
    \end{center}
\unboldmath
\vspace{0.2cm}
\begin{center}
{  {\large Xavier Calmet}\footnote{x.calmet@sussex.ac.uk}}
 \end{center}
\begin{center}
{\sl Department of Physics and Astronomy, 
University of Sussex, Brighton, BN1 9QH, United Kingdom
}\end{center}
\vspace{5cm}
\begin{abstract}
\noindent
In this paper, we consider a generic hidden sector which interacts only gravitationally with Standard Model particles. We show that quantum gravity leads to operators which can be probed with fifth force type experiments. The E\"ot-Wash torsion pendulum experiment implies that the masses of any scalar field or any massive spin-2 field that couples with the usual gravitational strength to the energy-momentum tensor of the Standard Model must be larger than $10^{-3}$eV. This has interesting consequences for models of dark matter which posit very light scalar fields.  Dark  matter must be heavier than $10^{-3}$eV if it is a scalar field or a massive spin-2 field. 
\end{abstract}  
\vspace{5cm}
\end{titlepage}



\newpage

\section{Introduction}

The possibility of a hidden sector of particles beyond the Standard Model is a very interesting speculation which could potentially account for the missing dark matter that we know interacts very weakly with the particles of the Standard Model. From a theoretical point of view, hidden sectors are expected in many extensions of the Standard Model, ranging from grand unified theories to models that incorporate gravity such as string theory.  

In this paper we make a very simple point. As gravity is universal, any hidden sector must interact to some level with the fields of the Standard Model. We will argue that gravitational interactions must lead to effective operators that connect Standard Model fields to fields of the hidden sector.  We show that these interactions are sufficient to set a limit on the mass of any scalar or tensor field of a hidden sector.

In this paper, we will consider a generic hidden sector that consists of scalar fields $\phi_i$, pseudo-scalars fields $a_i$, vector bosons $B_{\mu,i}$ (with a corresponding field strength $B_{\mu\nu,i}$ which are anti-symmetrical in their two indices), spinors $\chi_i$ which could be Dirac or Majorana fields, and spin-2 fields $t_{\mu\nu,i}$ (which are symmetrical in their two indices). We will show that gravity will induce interactions between any of these fields and the Standard Model particles. Some of these interactions could be searched for with tension pendulum experiments such as the E\"ot-Wash experiment \cite{Kapner:2006si,Hoyle:2004cw,Adelberger:2006dh} 

In particular, we shall use the fact that the E\"ot-Wash experiment sets a strong bound on the mass $m$ of a new light field which leads to a fifth force type of interaction between Standard Model particles.  This experiment probes deviation from the $1/r$ Newton potential. The deviation is parametrized as $V(r)=-G \frac{m_1 m_2}{r} (1+\omega \exp(- r/\lambda))$. For $\omega=1$ which corresponds to a fifth force with a coupling constant of gravitational strength, the non-observation of a deviation from the $1/r$ potential leads to a bound on $\lambda$ which must be smaller than $0.03$cm or equivalently on the mass of the new particle responsible for the mediation of the fifth force $m=1/\lambda>4 \times 10^{-3}$eV $\sim 10^{-3}$eV. 

We will discuss both perturbative and nonperturbative quantum gravitational effects using a very conservative approach. The perturbative quantum gravitational calculations are based on the effective field theory approach to quantum gravity which enables model independent calculations. In the case of nonperturbative quantum gravitational effects, we simply classify operators in terms of symmetries.

This paper is organized as follows, we first discuss  perturbative quantum gravitational effects showing that they are too weak to lead to any bound on a hypothetical hidden sector. We then turn our attention to nonpertubative effects and show that quantum gravity leads to an interesting bound on the masses of any scalar field or massive spin-2 field belonging to this hidden sector. Finally we conclude.

\section{Perturbative quantum gravity}

In this section we work within the framework of the effective quantum gravitational action \cite{Weinberg,Bar1984,Bar1985,Bar1987,Bar1990,Buchbinder:1992rb,Donoghue:1994dn,Calmet:2018elv} which is  given by  
\begin{eqnarray}
\Gamma[g] = \Gamma_{\rm L}[g] + \Gamma_{\rm NL}[g],
\end{eqnarray}
where the local part of the action is given by
\begin{align}
	\Gamma_{\rm L} &= \int d^4 x \sqrt{g} \left[ \frac{1}{16\pi G_N} R +L_{M}+ c_1(\mu) R^2 + c_2(\mu) R_{\mu\nu} R^{\mu\nu} + c_3(\mu) R_{\mu\nu\alpha\beta} R^{\mu\nu\alpha\beta} \right] \ ,
	\end{align}
	where $L_{M}$ is the matter sector Lagrangian containing all of the matter fields and the non-local part of the action by
	\begin{align}
	\Gamma_{\rm NL} &= - \int d^4 x \sqrt{g} \left[ \alpha R \log{\left(\frac{\Box}{\mu^2}\right)} R 
	+ \beta R_{\mu\nu} \log{\left(\frac{\Box}{\mu^2}\right)} R^{\mu\nu} + \gamma R_{\mu\nu\alpha\beta} \log{\left(\frac{\Box}{\mu^2}\right)} R^{\mu\nu\alpha\beta} \right]
	\ ,
	\label{eq:NLterms}
\end{align}
where $\mu$ is a renormalization scale, see e.g. \cite{Donoghue:2017pgk}.
This effective action is obtained by integrating out the one-loop quantum fluctuations of the graviton. Focussing on the local part of the action and mapping it to the Einstein frame, one obtains after linearization:
\begin{eqnarray}
S=\int d^4 x \left[\left (- \frac{1}{2} h_{\mu\nu} \Box h^{\mu\nu}
 +\frac{1}{2} h_{\mu}^{\ \mu} \Box h_{\nu}^{\ \nu}  -h^{\mu\nu} \partial_\mu \partial_\nu h_{\alpha}^{\ \alpha}+ h^{\mu\nu} \partial_\rho \partial_\nu h^{\rho}_{\ \mu}\right) \right. \\ 
\left. \nonumber -\left ( -\frac{1}{2} k_{\mu\nu} \Box k^{\mu\nu}
 +\frac{1}{2} k_{\mu}^{\ \mu} \Box k_{\nu}^{\ \nu}  -k^{\mu\nu} \partial_\mu \partial_\nu k_{\alpha}^{\ \alpha}+ k^{\mu\nu} \partial_\rho \partial_\nu k^{\rho}_{\ \mu}
 \right.  \right.  \\ \left. \left. \nonumber
 -\frac{M_2^2}{2} \left (k_{\mu\nu}k^{\mu\nu} - k_{\alpha}^{\ \alpha} k_{\beta}^{\ \beta} \right )
 \right)  \right.
  \\
  \nonumber
 \left. + \frac{1}{2} \partial_\mu \sigma  \partial^\mu \sigma
  - \frac{M_0^2}{2} \sigma^2 - \sqrt{8 \pi G_N} (h_{\mu\nu}+k_{\mu\nu}+\frac{1}{\sqrt{3}} \sigma \eta_{\mu\nu})T^{\mu\nu} 
  \right ],
\end{eqnarray}
where the masses of the spin-2 $k_{\mu\nu}$ and spin-0 $\sigma$  fields are given by
\begin{align}
M_2^2 =\frac{ M_P^2}{-2 c_2 }, \quad M_0^2 =\frac{M_P^2}{4(3 c_1 + c_2 )} \ \ .
\end{align}
Furthermore, $T^{\mu\nu}$ is the sum of the energy momentum tensor of the Standard Model $T^{\mu\nu}_{SM}$ and that of the hidden sector $T^{\mu\nu}_H$. If the fields $k_{\mu\nu}$ and   $\sigma$ are heavy, they will lead to contact interactions between fields of the Standard Model and that of the hidden sector even if we do not introduce couplings between these sectors at tree level. For example, in the case of the spin-0 field, perturbative quantum gravity leads to operators of the type
\begin{align}
O_{pert}= \frac{8 \pi}{M_P^2 M_0^2} T_{SM} T_{H}
\end{align}
These are dimension 8 operators and they are suppressed by four powers of the Planck mass. Bounds from the torsion pendulum E\" ot-Wash experiment on $3 c_1 + c_2$ are rather weak, all that is known is that $3 c_1 + c_2\le10^{61}$, but even in the most optimistic case, these operators are essentially irrelevant for any low energy experiment. 

If the only coupling between the Standard Model and the hidden sector is via gravitational interactions, it is easy to see that graviton loops will also induce operators that are suppressed by at least four powers of the Planck mass. Furthermore, they will be suppressed by powers of $1/(16 \pi^2)$ depending on the loop order. Perturbative quantum gravity effects can only lead to minute interactions between the Standard Model fields and those of the hidden sector. As these interactions are described by operators of dimension 8, they cannot be relevant to low energy experiments. We shall see very soon that nonperturbative effects can lead to much stronger interactions.

\section{Nonperturbative quantum gravity}

While, as we have just seen, perturbative quantum gravitational effects will at best generate dimension 8 operators linking the Standard Model sector to the hidden sector if the coupling between the two sectors is only gravitational, nonperturbative effects could generate operators of lesser dimensions which could thus be of relevance to low energy physics experiments.

Nonperturbative effects in quantum gravity are still poorly understood, but given our experience of strongly coupled quantum field theoretical systems such as quantum chromodynamics, we are able to make a few model independent predictions. If quantum gravity effects such as quantum black holes or wormholes  lead to new interactions between the Standard Model fields and those of a hypothetical hidden sector, these effects must vanish in the limit of $M_P$ going to infinity  \cite{Holman:1992us,Barr:1992qq,Calmet:2009uz,Calmet:2014lga}. This implies that dimension four operators must be exponentially suppressed and proportional to $e^{-M_P}$. Higher dimensional operators must be suppressed by powers of the Planck scale. Note that there is no loop suppression factor as these operators are expected to be generated via nonperturbative effects. The Wilson coefficients of these operators are thus expected to be of order one.

It has been argued that nonperturbative effects corresponding to virtual quantum black holes or wormholes could even lead to operators that may not respect global symmetries \cite{Holman:1992us,Barr:1992qq,Kallosh:1995hi}. In that case, it has been argued but not proven, that the  Wilson coefficients 
could be further suppressed by a factor $e^{-S}$ \cite{Kallosh:1995hi}. Here, $S$ is the action of gravitational instanton responsible for the symmetry breaking. We will focus on operators that do not violate any global symmetry, there is thus no ground to expect such a suppression.  For example, quantum black hole production during the scattering of high energetic particles has been shown not to be exponentially suppressed \cite{Hsu:2002bd}. There is thus no obvious reason to expect that the Wilson coefficients $c_i$ of the operators considered here should be suppressed and they are expected to be of order one as explained previously. As gravity is universal, we expect the Wilson coefficients to be identical for all fields. 

We are interested in operators that connect fields from the Standard Model with those of the hidden sector creating a portal between the two sectors. These interactions can be organized according to the dimension of the effective operators generated by gravitational nonperturbative effects. Obviously such operators must be compatible with the space-time symmetries, such as Lorentz invariance, and the gauge symmetries of the Standard Model. 

These few facts enable us to classify all possible operators given the particle content of the Standard Model and that of the hidden sector. We use $H$ to denote the SU(2) Higgs doublet, $F^{\mu\nu}_Y$ to denote the field strength of the hyperphoton $A_\mu$.  The field strength of the SU(2) gauge bosons in denoted by $F_{\mu\nu, 2}$ while that of QCD is denoted by $G_{\mu\nu}$. We use $\Psi_L$ and $\Psi_R$ to denote respectively the left and right-handed fermion fields of the Standard Model. We now present the lowest order operators which connect the Standard Model to a hidden sector composed of scalar fields $\phi_i$, pseudo-scalar fields $a_i$, fermions $\chi_i$, massive vector fields $B_{\mu,i}$ and massive spin-2 tensor fields $t_{\mu\nu,i}$.
\begin{itemize}
\item Hidden sector scalar fields.

The Higgs portal coupling enables a coupling to scalar fields $\phi_i$ via a dimension four operator, however such operators must vanish in the limit where $M_P \to \infty$. We thus have
\begin{eqnarray}
 O_1&=&c_s e^{-M_P/\mu} H^\dagger H \phi_i \phi_j,
 \end{eqnarray}
 where $\mu$ is some low energy scale. We thus expect this operator to be extremely suppressed and irrelevant for all practical purposed.
 
Dimension five operators can be generated via scalar fields of the hidden sector coupling either to the Lagrangian of the Standard Model $L_{SM}$ or to the trace of the energy-momentum tensor $T_{SM}^{\mu\nu}$
\begin{eqnarray}
 O_2&=&  c_s \frac{\sqrt{4 \pi G_N}}{4}  \phi_i L_{SM}\\
 O_3&=&  c_s \frac{\sqrt{4 \pi G_N}}{4}  \phi_i T_{SM}.
 \end{eqnarray}
Note that the trace of the gauge sector of the tree level energy-momentum tensor vanishes. The most important operator is 
\begin{eqnarray}
\frac{\sqrt{4 \pi G_N}}{4} c_s \phi_i G_{\alpha\beta}  G^{\alpha\beta}
 \end{eqnarray}
as it leads to a fifth force between regular matter particles. The E\"ot-Wash experiment implies that the mass of $\phi$ must be larger than $10^{-3}$eV. As emphasized already, these operators do not violate any symmetry. As explained above, there is no reason to expect an exponential suppression of the Wilson coefficient $c_s$ and it should be of order one.

\item Hidden sector pseudo-scalar fields

Here we focus on the couplings of pseudo-scalar fields $a_i$ (axion like particles) to the Standard Model:
\begin{eqnarray}
O_4&=& \frac{\sqrt{4 \pi G_N}}{4}  (c_{ps} a_i F_{\alpha\beta, Y}  \tilde F^{\alpha\beta}_Y+c_{ps}  a_i   F_{\alpha\beta, 2} \tilde F^{\alpha\beta}_2+c_{ps} a_i F_{\alpha\beta, G}\tilde  F^{\alpha\beta}_G) \\
O_5&=& \frac{ \pi G_N}{4} c_{ps} \partial_\mu a_i  \bar \Psi_R \gamma^\mu \gamma^5 H \Psi_L
\\
O_6&=& \frac{ \pi G_N}{4} c_{ps} \partial_\mu a_i  \partial^\mu a_i   H^\dagger H.
\end{eqnarray}
There are only weak bounds on axion like particles coupling to $F \tilde F$ if the suppression scale is of the order of the Planck mass as assumed here. The superradiance instability of astrophysical black holes  excludes masses of axion like particles in the range $6\times 10^{-13}$eV to $10^{-11}$eV  \cite{Cardoso:2018tly}.

\item Hidden sector fermions  
\begin{eqnarray}
O_7&=& \frac{\sqrt{4 \pi G_N}}{4} c_\chi \bar \chi_i \sigma_{\mu\nu}  \chi_i F^{\mu\nu}_Y\\ 
O_8&=& \frac{ \pi G_N}{4} c_\chi  \bar \chi_i  \chi_i \bar \Psi_R H \Psi_L
\end{eqnarray}
The mass of such fermions is unconstrained as any cross-section with regular matter is of the order of $1/M_P^2$. Note that they do not lead to a new force.

\item Hidden sector vector fields
\begin{eqnarray}
 O_9&=&c_{B}e^{-M_P/\mu} F^{\mu\nu}_Y B_{\mu\nu,i},
 \end{eqnarray}
where $\mu$ is some low energy scalar of the order of the mass of the $B^{\mu}_i$ boson. We thus expect the coupling between the two U(1) sectors to be exponentially suppressed if it is of gravitational origin. There are only weak bounds on the masses of extra U(1) massive photons from 
the superradiant instability of astrophysical black holes. Masses in the range $10^{-13}$eV to  $3\times 10^{-13}$eV are disfavored \cite{Cardoso:2018tly}.  

If the vector field does not carry a new charge, we can add $ \frac{ \pi G_N}{4} \bar \Psi_R \gamma_\mu H \Psi_L B^{\mu}$ to our list. This operator would mediate a long range force if the mass of $B^\mu$ is light enough albeit it would be suppressed by a factor $v/M_P$ where $v=246$ GeV is the Higgs boson's vacuum expectation. Such a weak fifth force is not constrained by torsion pendulum experiments. A pseudo-vector coupling of the type $\frac{ \pi G_N}{4} \bar \Psi_R \gamma_\mu \gamma_5 H \Psi_L B^{\mu}_i$ would have the same chiral suppression.
 
\item Hidden sector spin-2 massive fields 
\begin{eqnarray}
O_{10}&=&\frac{\sqrt{4 \pi G_N}}{4}  \Big (c_{t} t_{i,\mu\nu} (F_Y^{\mu\alpha}F^{\nu}_{\alpha,Y}-\frac{1}{4} \eta^{\mu\nu} F_{\alpha\beta,Y}F_Y^{\alpha\beta})\\  \nonumber && +c_t t_{i,\mu\nu} (F_2^{\mu\alpha}F^{\nu}_{\alpha,2}-\frac{1}{4} \eta^{\mu\nu} F_{\alpha\beta,2}F^{\alpha\beta}_2) \\  \nonumber && +c_{t} t_{i,\mu\nu} (G^{\mu\alpha}G^{\nu}_{\alpha}-\frac{1}{4} \eta^{\mu\nu} G_{\alpha\beta}G^{\alpha\beta}) \Big)\\
O_{11}&=&\frac{\sqrt{4 \pi G_N}}{4} c_t t_{i,\mu\nu} \left (\left ( -\frac{1}{2} \partial_\alpha H^\dagger \partial^\alpha H- V(H^\dagger H) \right) g^{\mu\nu} +\partial^\mu H^\dagger \partial^\nu H \right)\\
O_{12}&=&\frac{\sqrt{4 \pi G_N}}{4} c_{t} t_{i,\mu\nu} \left (
i  \bar \Psi_R\gamma^\mu \partial^\nu \Psi_{L} - \eta^{\mu\nu} \bar \Psi_R (i \gamma^\alpha \partial_\alpha-m) \Psi_{L} \right).
\end{eqnarray}
As for the scalar field, these operators lead to a fifth force and the E\"ot-Wash experiment implies that the masses of massive spin-2 fields $t_{i,\mu\nu}$ must be larger than $10^{-3}$eV. As in the case of the scalar fields, these operators do not violate any symmetry. There is no reason to expect that the Wilson coefficients $c_{t}$ are not of order one.
\end{itemize}

\section{Discussion and conclusions}
Gravity is universal and will thus generate interactions between the Standard Model and any hypothetical hidden sector. Remarkably and despite its weakness, gravity has very deep implications for fields of any hidden sector. In particular, we have shown that quantum gravity will generate an interaction between regular matter and any field of a hidden sector. In the case of scalars and tensor fields from a hidden sector, the E\"ot-Wash torsion pendulum experiment implies that the masses of these boson fields are heavier than $10^{-3}$eV independently of whether they are a form of dark matter or not.  On the other hand, the masses of massive vector fields, pseudo-scalars and fermions from a hidden sector remain unconstrained.  Quantum gravity will not generate a sizable interaction between U(1) sectors as such interactions would be represented by dimension four operators which must be exponentially suppressed.

Gravity is expected to be the weakest force in nature \cite{ArkaniHamed:2006dz}.  However, even if we were willing to accept interactions between the Standard Model fields and a hidden sector that are weaker than gravity, gravity would still generate measurable interactions between scalar fields and massive spin-2 fields from a hidden sector and regular matter if these bosons are very light. This has remarkable implications for the dark matter sector if it is composed by such particles. These dark matter fields must be heavier than $10^{-3}$eV. 

\bigskip{}

{\it Acknowledgments:}
This work is supported in part  by the Science and Technology Facilities Council (grant number ST/P000819/1).


\bigskip{}


\baselineskip=1.6pt 

\begin{thebibliography}{10}

\bibitem{Kapner:2006si} 
  D.~J.~Kapner, T.~S.~Cook, E.~G.~Adelberger, J.~H.~Gundlach, B.~R.~Heckel, C.~D.~Hoyle and H.~E.~Swanson,
  Phys.\ Rev.\ Lett.\  {\bf 98}, 021101 (2007)
  doi:10.1103/PhysRevLett.98.021101
  [hep-ph/0611184].

\bibitem{Hoyle:2004cw} 
  C.~D.~Hoyle, D.~J.~Kapner, B.~R.~Heckel, E.~G.~Adelberger, J.~H.~Gundlach, U.~Schmidt and H.~E.~Swanson,
  Phys.\ Rev.\ D {\bf 70}, 042004 (2004)
  doi:10.1103/PhysRevD.70.042004
  [hep-ph/0405262].

\bibitem{Adelberger:2006dh} 
  E.~G.~Adelberger, B.~R.~Heckel, S.~A.~Hoedl, C.~D.~Hoyle, D.~J.~Kapner and A.~Upadhye,
  Phys.\ Rev.\ Lett.\  {\bf 98}, 131104 (2007)
  doi:10.1103/PhysRevLett.98.131104
  [hep-ph/0611223].


\bibitem{Weinberg}
S.~Weinberg, in ``General Relativity. An Einstein Centenary Survey'', ed. by S.W. Hawking and W. Israel, (Cambridge University Press, Cambridge, 1979) p. 790.

  
   \bibitem{Bar1984} 
  A.~O.~Barvinsky and G.~A.~Vilkovisky,
  ``The Generalized Schwinger-de Witt Technique And The Unique Effective Action In Quantum Gravity,''
  Phys.\ Lett.\  {\bf 131B}, 313 (1983).
  
  \bibitem{Bar1985} 
  A.~O.~Barvinsky and G.~A.~Vilkovisky,
  Phys.\ Rept.\  {\bf 119}, 1 (1985).
  doi:10.1016/0370-1573(85)90148-6
  
   \bibitem{Bar1987} 
  A.~O.~Barvinsky and G.~A.~Vilkovisky,
  ``Beyond the Schwinger-Dewitt Technique: Converting Loops Into Trees and In-In Currents,''
  Nucl.\ Phys.\ B {\bf 282}, 163 (1987).
  
  \bibitem{Bar1990} 
  A.~O.~Barvinsky and G.~A.~Vilkovisky,
  ``Covariant perturbation theory. 2: Second order in the curvature. General algorithms,''
  Nucl.\ Phys.\ B {\bf 333}, 471 (1990).

\bibitem{Buchbinder:1992rb} 
  I.~L.~Buchbinder, S.~D.~Odintsov and I.~L.~Shapiro,
  Bristol, UK: IOP (1992) 413 p
  
\bibitem{Donoghue:1994dn} 
  J.~F.~Donoghue,
  Phys.\ Rev.\ D {\bf 50}, 3874 (1994)
  doi:10.1103/PhysRevD.50.3874
  [gr-qc/9405057].
  
  
    
\bibitem{Calmet:2018elv} 
  X.~Calmet,
  Phys.\ Lett.\ B {\bf 787}, 36 (2018)
  doi:10.1016/j.physletb.2018.10.040
  [arXiv:1810.09719 [hep-th]].

\bibitem{Donoghue:2017pgk} 
  J.~F.~Donoghue, M.~M.~Ivanov and A.~Shkerin,
  ``EPFL Lectures on General Relativity as a Quantum Field Theory,''
  arXiv:1702.00319 [hep-th].

\bibitem{Holman:1992us} 
  R.~Holman, S.~D.~H.~Hsu, T.~W.~Kephart, E.~W.~Kolb, R.~Watkins and L.~M.~Widrow,
  Phys.\ Lett.\ B {\bf 282}, 132 (1992)
  [hep-ph/9203206].
  
\bibitem{Barr:1992qq} 
  S.~M.~Barr and D.~Seckel,
  Phys.\ Rev.\ D {\bf 46}, 539 (1992).
  doi:10.1103/PhysRevD.46.539


\bibitem{Kallosh:1995hi} 
  R.~Kallosh, A.~D.~Linde, D.~A.~Linde and L.~Susskind,
  Phys.\ Rev.\ D {\bf 52}, 912 (1995)
  doi:10.1103/PhysRevD.52.912
  [hep-th/9502069].

\bibitem{Calmet:2009uz} 
  X.~Calmet and S.~K.~Majee,
  Phys.\ Lett.\ B {\bf 679}, 267 (2009)
  doi:10.1016/j.physletb.2009.07.049
  [arXiv:0905.0956 [hep-ph]].
  
\bibitem{Calmet:2014lga} 
  X.~Calmet and V.~Sanz,
  Phys.\ Lett.\ B {\bf 737}, 12 (2014)
  doi:10.1016/j.physletb.2014.08.022
  [arXiv:1403.5100 [hep-ph]].

\bibitem{Hsu:2002bd} 
  S.~D.~H.~Hsu,
  Phys.\ Lett.\ B {\bf 555}, 92 (2003)
  doi:10.1016/S0370-2693(03)00012-1
  [hep-ph/0203154].
  
\bibitem{Cardoso:2018tly} 
  V.~Cardoso, Ó.~J.~C.~Dias, G.~S.~Hartnett, M.~Middleton, P.~Pani and J.~E.~Santos,
  JCAP {\bf 1803}, no. 03, 043 (2018)
  doi:10.1088/1475-7516/2018/03/043
  [arXiv:1801.01420 [gr-qc]].
  

\bibitem{ArkaniHamed:2006dz} 
  N.~Arkani-Hamed, L.~Motl, A.~Nicolis and C.~Vafa,
  JHEP {\bf 0706}, 060 (2007)
  doi:10.1088/1126-6708/2007/06/060
  [hep-th/0601001].
  
  


\end{thebibliography}
\end{document}